# Towards Advanced Chiral Sensors: Enhanced Helicity-Dependent Photocurrent in Ultrathin Topological Insulator Films

Mohammad Shafiei,[1] Sahar Safavi Moayeri,[1] and Milorad V. Milošević[1,2, *]

[1] *Department of Physics, University of Antwerp, Groenenborgerlaan 171, B-2020 Antwerp, Belgium*
[2] *NANOlight Center of Excellence, University of Antwerp, B-2020 Antwerp, Belgium*
(Dated: November 18, 2024)

Chirality, a fundamental property of asymmetric structures, plays a crucial role in pharmaceutical, biological and chemical systems, offering a powerful tool for screening organic compounds. While the conventional optical chirality detectors are often bulky and involuted, the topological insulators (TIs) offer a promising platform for developing compact yet sensitive devices - owing to their inherent chirality. However, the complex interplay of photoresponses in TIs can limit the ultimate accuracy of chirality detection. Therefore, we here analyze the underlying mechanisms governing the photoresponses in TIs and reveal strategies to enhance the helicity-dependent photocurrent (HDPC). By attentively analyzing the symmetries and behavior of competing photoresponses, we show that it is possible to effectively eliminate unwanted contributions and isolate the HDPC. Moreover, we reveal that HDPC is strongly amplified in ultrathin TI films, and can be further enhanced by optimizing the illumination parameters, sensor strain and/or back gating. Our findings thereby provide a roadmap for design and optimization of miniaturized, high-performance TI chirality detectors, with potential to revolutionize chiral analysis in biomedical and material sciences.

## INTRODUCTION

Chirality emerges in systems that lack both a plane and a center of symmetry, resulting in two non-superimposable mirror-image system forms known as enantiomers [1, 2]. This phenomenon is pivotal in comprehending the processes occurring within living organisms, which predominantly function in an enantioselective manner [3, 4]. In physics language, a chiral system exists in two distinct enantiomeric states that can be interconverted by parity but not by time reversal [5]. Given the prevalence of chiral bioactive substances in nature, chirality holds immense significance in medical, pharmaceutical, clinical, and biological domains [4]. Initially termed "dissymmetry" by its discoverer, Louis Pasteur, the concept of biological chirality, despite being observed nearly 150 years ago, remains one of the most intriguing challenges in natural sciences [5]. Phenomenologically, biological chirality implies that chiral biomolecules within all living organisms exclusively appear as one of their enantiomers, or exhibit a significant excess of the preferred stereoisomer. Intriguingly, all living beings adhere to the same selection rules, favoring the same enantiomer of chiral biomolecules [5].

By discerning molecular chirality and elucidating the mechanisms underlying chiral biological phenomena, researchers have made significant strides in fields such as drug delivery, antimicrobial and antiviral activities, and biological signal detection [6–8]. Exploiting chirality offers the potential to enhance drug delivery systems, tumor markers, and biosensors, among other biomaterial-based technologies [9, 10]. Figure 1 provides a schematic representation of chirality, accompanied by examples of its biomedical effects and applications.

For the detection of chirality, various chiral sensors have been proposed and constructed using different methods, including chiral electrochemical sensors [11], chiral sensors based on quartz crystal microbalance [12], and optical chiral sensors [4]. Among these, optical chiral sensors have gained prominence due to their advantages such as rapid analysis, ease of handling, low cost, high sensitivity, and adaptability to enantiomers. These sensors exploit the distinct optical response exhibited by enantiomers with different chiralities, enabling the selective detection and identification of chiral species [4]. However, these conventional optical chirality detectors mostly employ semiconductors that lack intrinsic chiral response. This necessitates the inclusion of additional components, such as rotating optomechanical elements, polarizing beam splitters, and waveplates, resulting in relatively bulky and complex setups [13]. Consequently, there is a pressing need for more compact and efficient solutions. In that respect, topological insulators (TIs) have recently emerged as promising candidates for addressing these challenges [13, 14]. TIs are characterized by a bulk insulating state and conducting edge or surface states [15], with the $Bi_2Se_3$ family of materials being the best known host of such a quantum state. In topological insulators, the spin of surface state electrons is locked to their momentum, that inherently results in a chiral response. This allows for the induction of spin polarization by applying a current and vice versa [16, 17]. By injecting spin polarization into these materials using polarized light gives rise to a photogalvanic effect, and a directional DC photocurrent is generated, enabling highly accurate optical chirality detection [13]. Notably, since the photocurrent originates from the surface states, chirality detection can extend to the infrared range, enabling direct measurement of chiral thermal radiation emitted by these materials [18, 19].

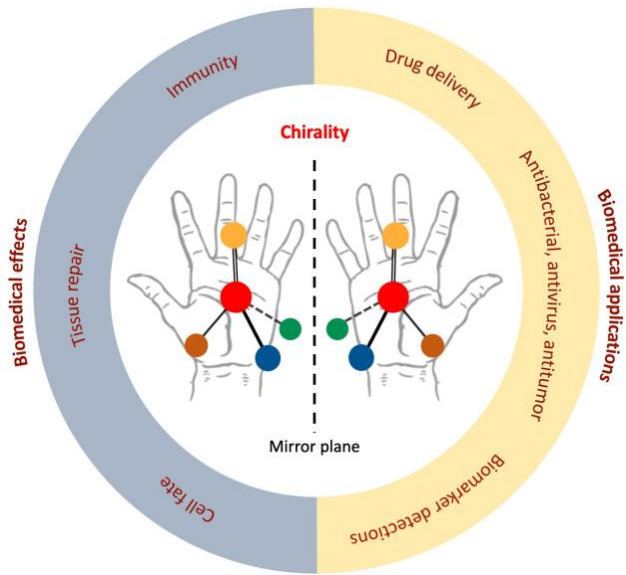

FIG. 1. Schematic representation of chirality, highlighting its biomedical effects and applications. A chiral system exhibits two distinct enantiomeric states, each a non-superimposable mirror image of the other. Enantiomers can be interconverted through parity operations but not time reversal.

Therefore, TIs offer a promising avenue for optical chirality detection, also due to their ease of integration and fabrication at small scales. However, other challenges emerge, such as the complexity of their photoresponse. Namely, the coexistence of circular photogalvanic effect (CPGE), linear photogalvanic effect (LPGE), and linear photon-drag effect (LPDE) in these materials complicates the direct chirality detection of elliptically polarized light [17, 20, 21]. Consequently, it became critical to develop methods for isolating and identifying the undesired linear photogalvanic and photon-drag effects from the desired circular photogalvanic response. Previous studies by Huang *et al.* [13] and Chen *et al.* [14] have demonstrated that applying a combination of gate and bias voltage enables the discrimination of helicitydependent photocurrent from polarization-dependent photocurrents. Specifically, they have shown that differential photocurrent calculations for varied bias voltages can isolate the circular polarization component, as it is independent of linear polarization. While this approach is promising for single-device chirality detection, the uncertainty in photoconductance, which directly correlates with the chiral sensitivity of the detector, remains a significant limitation. Reported values of approximately 5.6% [13] indicate that only chirality associated with circularly polarized light components exceeding this threshold can be reliably detected. In comparison, conventional metamaterial-based chiral detectors typically exhibit a chiral sensitivity around 3% [22]. Moreover, the needed application of a gate voltage can induce a shift in the chemical potential towards the bulk bands, which can increase the scattering rate out of the topological surface states and diminish the circular photoresponse. Thus, although such TI-based transistors enable the realization of compact, low-fabrication-effort chirality detectors, their sensitivity falls short of conventional metamaterial-based devices. Consequently, developing strategies to enhance the chiral sensitivity of TI-based optical chiral biosensors remains a critical research objective. In that respect, one seeks control and enhancement of CPGE in TIs, to substantially improve the helicity-dependent photocurrent (HDPC) relative to the total polarization-dependent photocurrent.

In this paper, we present a comprehensive analysis of the photocurrent responses in TIs, focusing on the CPGE, LPGE, and LPDE. By investigating the dependence of these responses and associated effects on the light irradiation angle ($\theta$, cf. Figure 2), based on their nature and available experimental data, we identify a method to isolate the HDPC, which is crucial for optical chirality detection. Leveraging the distinct physical origins of CPGE, LPGE, and LPDE, we show that this approach enables the identification of the HDPC without the need for external bias voltages, facilitating the realization of compact, single-device optical chirality detectors. Furthermore, we show that using ultrathin TI films for the same purpose not only suppresses bulk contributions but also significantly enhances the HDPC, thereby further improving the chiral sensitivity. Our findings highlight the importance of engineering both the TI (thickness, band structure) and the used incident light (angle, frequency, intensity, and duration) to optimize the HDPC and achieve a superior signal-to-noise ratio. Moreover, we demonstrate that applying a gate voltage, and/or compressive biaxial strain to ultrathin TI films can additionally enhance the helicity-dependent optical response, making such films a prime platform for optical chirality detectors. These findings resolve the key challenges in the development of TI-based optical chirality detectors and pave the way for compact, high-performance devices with broad applications in biomedical diagnostics and therapeutics.

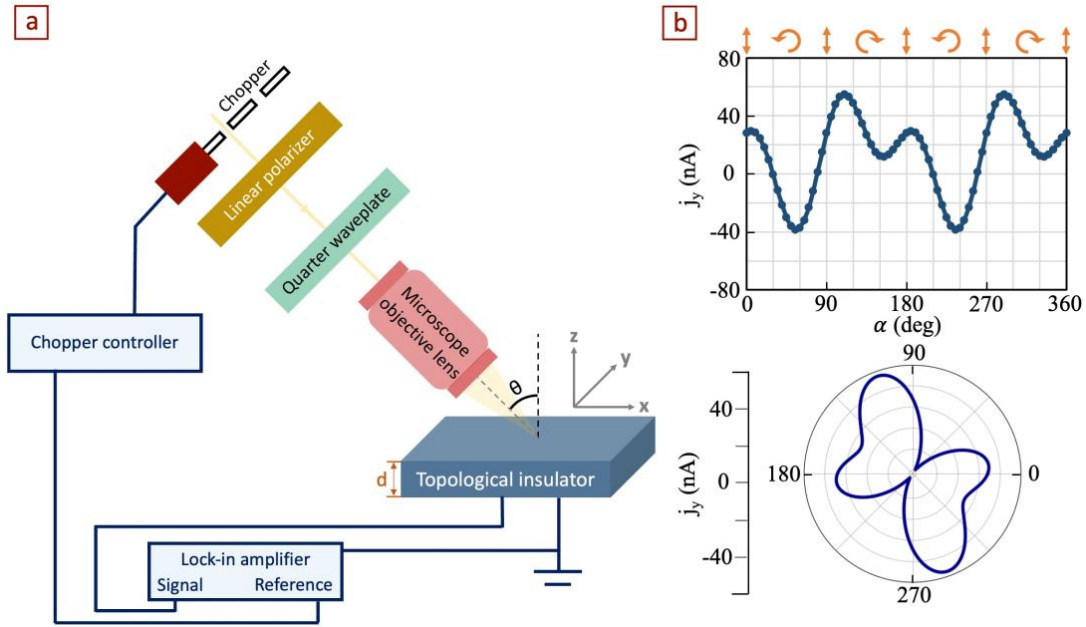

FIG. 2. (a) Schematic representation of the setup to detect the photocurrent generated by circularly polarized light under oblique incidence. A continuous-wave laser beam is modulated using a chopper and linearly polarized by a polarizer, while a quarter-wave plate (QWP) is employed to control the polarization of light. The beam is focused onto the sample using a microscope objective lens and the photocurrent generated in the sample is measured using a lock-in amplifier. (b) Polarizationdependent photocurrent of the TI as a function of the polarization angle $\alpha$, as tuned by QWP. The parameters of Eq. (1) were set to $C = -28.5$, $L_1 = 21.1$, $L_2 = 14.8$, and $D = 13.2$ nA, as obtained in Ref. [13] by fitting experimental data.

## RESULTS AND DISCUSSION
### Methodological backbone

Topological insulators are prime candidates for intrinsic solid-state polarization detection owing to their unique electronic properties. Their metallic surface states exhibit linear dispersion, where spin-polarized carriers are locked perpendicular to their momentum [15, 23]. This spin-momentum locking, protected by time-reversal symmetry, is readily exploited in spintronics, optics, and optoelectronics [15, 24]. Optical methods are the primary approach to control spin-polarized currents in 3D TIs, with circularly polarized light inducing HDPC [17, 25]. However, the pertinent challenge lies in distinguishing these photocurrents from other effects like the LPGE, and LPDE, which hinder direct chirality detection [13, 14, 21].

To investigate the polarization-dependent photocurrent, we consider the sample illuminated with a laser beam that passed through a quarter waveplate (QWP), as depicted in Figure 2(a). In experiments, the laser beam is modulated by a chopper, passed through a linear polarizer and a QWP, and then focused onto the sample using a microscope objective lens. The QWP periodically alters the polarization state from linearly polarized (polarization angle $\alpha = 0$) to left circular ($\alpha = \pi/4$), then back to linearly polarized ($\alpha = \pi/2$), then right circular ($\alpha = 3\pi/4$), and finally to linearly polarized ($\alpha = \pi$), with a 180-degree periodicity. When the incident light is tilted in the $x - z$ plane, the $y$-component of the photocurrent exhibits a strong dependence on polarization and can be decomposed into four components:

$$j_y = C \sin 2\alpha + L_1 \sin 4\alpha + L_2 \cos 4\alpha + D. \tag{1}$$

Here, coefficient $C$ characterizes the (desired) HDPC, while $L_1$ and $L_2$ correspond to the helicity-independent photocurrents related to linear polarization. The term $D$ represents a polarization-independent component. Figure 2(b) also shows the polarization-dependent photocurrent of the TI as a function of the QWP-controlled polarization angle $\alpha$, obtained after fitting the experimental data with Eq. (1) using $C = -28.5$, $L_1 = 21.1$, $L_2 = 14.8$, and $D = 13.2$ nA [13]. This photocurrent density, $j_y$, contains contributions from CPGE, LPGE, and LPDE. To effectively employ TIs in optical chirality detection, it is crucial to isolate the HDPC arising from circular photocurrent from the other constituent components.

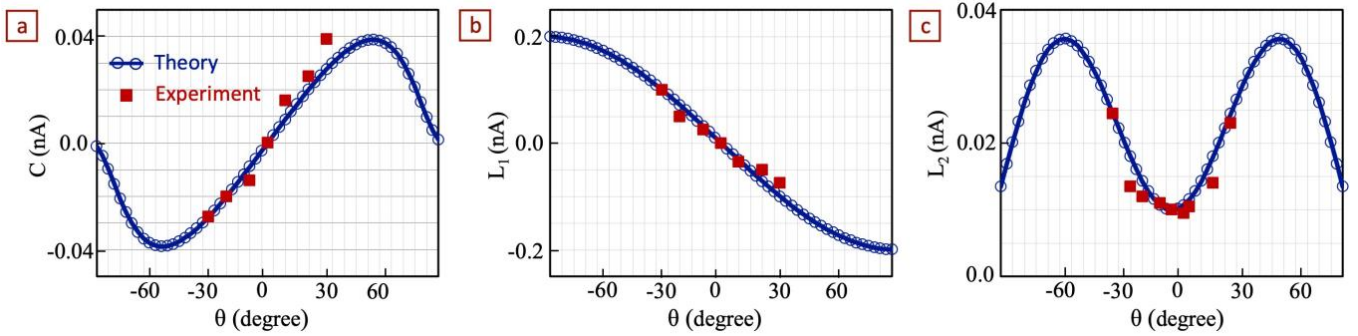

FIG. 3. Tilt-angle dependence of CPGE, LPGE, and LPDE in a TI film ($d \approx 10$ nm). Theoretical calculations (blue lines and points) are in good agreement with experimental data (red points, from Ref. [14]). The even parity of the $L_2$ component with respect to the angle of incidence $\theta$, enables the extraction of the pure LPDE contribution by calculating the photocurrent at $\theta$ and $-\theta$.

The CPGE in topological surface states is a nonlinear optical phenomenon stemming from the system's inversion asymmetry. This asymmetry results in an unequal excitation of states in $k$-space due to the transfer of angular momentum from incident photons to free carriers. Unlike the linear photogalvanic effect (LPGE) and linear photondrag effect (LPDE), CPGE is independent of linear momentum transfer and is constrained by crystal symmetry. Notably, the CPGE in surface states vanishes when the incident light is normal to the sample surface ($\theta = 0$), as it originates from the asymmetric optical excitation of the helical Dirac cone. The absence of inversion symmetry in the topological surface states, characterized by $C_{3v}$ symmetry, is a prerequisite for LPGE, while LPDE can occur in both surface and bulk states regardless of inversion symmetry. The polarization-independent $D$ term in Eq. (1), attributed to photothermoelectric and photovoltaic effects [26, 27], is highly sensitive to the specific location on the TI surface interacting with the incident light. To minimize the photothermoelectric effect, incident light is typically centered on the TI surface. As this term is polarization-independent, it does not influence the determination of the chirality of incident light, and we may neglect it in our analysis.

The CPGE current perpendicular to the plane of the incident light tilted under angle $\theta$ (x–z plane, cf. Figure 2), for the surface states of TI (with reduced symmetry of $C_{3v}$) can be expressed by the following equation [28, 29]:

$$j_{CPGE} = \frac{Y_{CPGE} \sin\theta \cos^2\theta}{n_\omega(\cos\theta + \sqrt{n_\omega^2 - \sin^2\theta})(n_\omega^2\cos^2\theta + \sqrt{n_\omega^2 - \sin^2\theta})}. \tag{2}$$

Here $n_\omega$ denotes the refractive index of the TI film, and the coefficients $Y_{CPGE}$ is directly related to the strength of spin-orbit coupling of TI. The LPGE and LPDE currents can be expressed as functions of the angle of incidence ($\theta$, cf. Figure 2) as [14, 20, 30]:

$$j_{LPGE} = Y_{LPGE} \sin\theta \, \sin 4\alpha, \tag{3}$$

and

$$j_{LPDE} = Y_{LPDE} \sin^2\theta \, \cos\theta \, \cos 4\alpha, \tag{4}$$

where the magnitudes of $Y_{LPGE}$ and $Y_{LPDE}$ are influenced by both the electric field amplitude of the incident light wave and the refractive index $n_\omega$ of the TI films [14].

With above understanding, we proceed to investigate the influence of the angle of incidence $\theta$ on both helicity-dependent and polarization-dependent photocurrents. The calculated currents associated with each component are summarized in Figure 3, where blue lines and points represent our theoretical calculations using Eqs. (2)-(4) and red points correspond to experimental data from Ref. [14] for a TI film of ~10 nm thickness ($n_\omega$ = 6 [31]), demonstrating very good functional agreement. Here, coefficients $Y_{CPGE}$, $Y_{LPGE}$ and $Y_{LPDE}$ were fixed such that theoretical results match the experimental ones for $\theta = -30°$. Notably, $L_2$ exhibits even parity with respect to $\theta$, enabling the isolation of the LPDE contribution by calculating the photocurrent at both $\theta$ and $-\theta$. The latter allows for the effective elimination of the LPDE component from the overall photocurrent response. The CPGE and LPGE components cannot be isolated from each other in this manner, but will exhibit different dependence on relevant parameters, as will be discussed further.

**Ultrathin TI films as advanced chirality detectors**

In what follows, we turn our attention to TI films in ultrathin regime, and their enhanced performance as optical chirality detectors. Namely, by reducing the TI thickness to a few nanometers, one can effectively mitigate bulk effects that interfere with polarization-dependent photoresponses, which should lead to a substantial increase in the circular photogalvanic current. Specifically, we thereby expect to significantly improve the ratio of HDPC to the sum of the absolute values of polarization-dependent photocurrents. This enhancement can be attributed to the suppression of the photon-drag effect related to the bulk [17, 33], which is associated with the transfer of linear momentum from incident photons to excited carriers. Furthermore, using ultrathin TI films allows for precise control of the chemical potential using a back-gate, enabling fine-tuning of the TI's electronic properties [33].

Figure 4(a) shows the temporal dynamics of ultrafast light-driven currents within low-dimensional TIs [32]. Photogalvanic currents are discernible for several picoseconds following ultrafast photoexcitation of the TI. On the other hand, photothermoelectric currents, arising from the thermalization of photoexcited charge carriers, exhibit lifetime ranging from hundreds of femtoseconds to several picoseconds [32]. To mitigate the influence of increasing photothermoelectric currents, we focused on a delay time ($\tau$) in 10-100 picosecond range for our calculations (cf. Figure 4(a)).

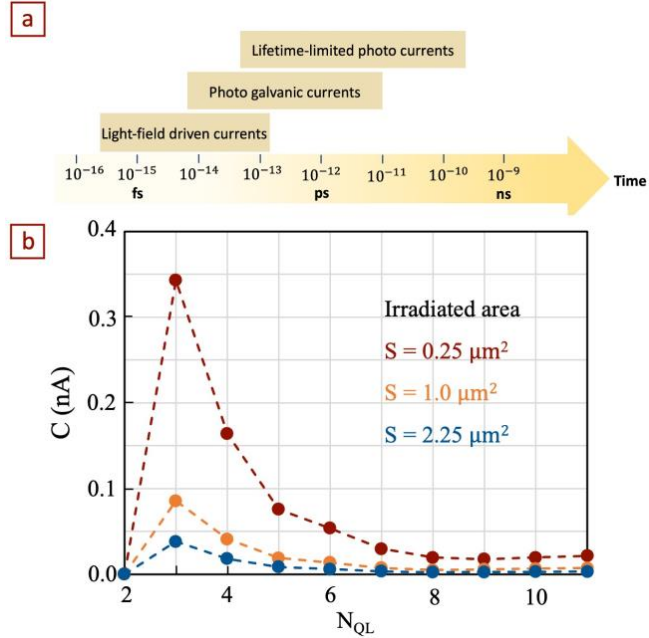

FIG. 4. (a) The temporal dynamics of ultrafast light-driven currents within low-dimensional TIs (after Ref. [32]). (b) Calculated photogalvanic current as a function of film thickness (expressed as number of quintuple layers $N_{QL}$) and irradiated area (focus of the light beam). The broken inversion symmetry for $N_{QL} > 2$ gives rise to a HDPC. Reducing the sample thickness enhances the photogalvanic current, reaching a maximum at 3QL thickness. Decreasing the irradiated area also enhances the photogalvanic current. For a 2QL sample, the presence of inversion symmetry nullifies the circular photogalvanic current. Here, the used parameters of incident light were $\theta = \pi/4$, $\omega$ =0.3 eV, $A_0$ =0.03 nm$^{-1}$, and $\tau$ =60 ps.

Bi$_2$Se$_3$-family TIs have van der Waals structure, with five atoms per unit cell and a hexagonal primitive cell with five atomic layers, known as the quintuple layer [34]. It is therefore readily feasible to fabricate crystalline ultrathin films of these materials, of thickness down to a single quintuple layer (QL, $d \approx 1$ nm), using molecular beam epitaxy (MBE) [35] and the vapor-liquid-solid method [36]. Through precise control of growth conditions, one can achieve layer-by-layer growth of TI ultrathin films [37]. In Figure 4(b), we therefore present the calculated photogalvanic current as a function of sample thickness $d$ (expressed as number of quintuple layers, N$_{QL}$) and the focus of the laser beam, denoted via the irradiated area, $S$. Experimental findings to date demonstrate that while the presence of a substrate breaks structural inversion symmetry, inducing a Rashba-type spin-splitting in ultrathin films, a 2QLthick sample retains inversion symmetry [37], leading to a vanishing circular photogalvanic current. However, for TIs thicker than 2QL, inversion symmetry is broken, resulting in nonzero HDPC. This symmetry breaking can arise from substrate-induced band bending or differing surface environments [36]. As shown in Figure 4(b), reducing the sample thickness, under conditions where structural inversion symmetry is broken, enhances the circular photogalvanic ($C$-)current. This helicity-dependent current reaches a maximum for the 3QL-thick sample. Moreover, decreasing the irradiated area for the same intensity of light increases the power density of illumination also increases the photogalvanic current. It is important to note that the detector cross-section should not be excessively small, in order not to hinder the ability to focus the laser solely onto the interior of the sensor and avoid edge effects and additional thermoelectric contributions. Our results shown in Figure 4(b) indicate that HDPC can be increased by more than an order of magnitude when thinning the TI film to a few quintuple layers, and that increasing the focus, i.e. reducing the irradiated area from 2.25 $\mu$m$^2$ to 0.25 $\mu$m$^2$, adds another order of magnitude to the HDPC signal. Such combined enhancement can thus radically improve the sensitivity and performance of chiral detectors.

As mentioned in the introduction, although thick TIs do hold potential for compact chirality detectors, experimental studies have revealed their relatively modest chiral sensitivity, approximately 5.6%, slightly lower than that of conventional metamaterial-based chiral detectors ($\approx$ 3%). The degree of chirality, or chiral sensitivity, is quantified by the ratio of the circular photogalvanic effect (CPGE) current to the total current, encompassing both linear and circular responses. As discussed, the $L_2$ (LPDE) current of the linear response (Eq. (4)) can be eliminated using its symmetry with respect to the tilt angle $\theta$. The other linear current, $L_1$ (LPGE), is expected to decrease with decreasing thickness of the TI film. Namely, the experimental data of Ref. [14] demonstrate that, when reducing the thickness from 20QL to 10QL, $L_1$ current reduces threefold (when averaged over all $\theta$). Since we find that ultrathin TI films exhibit a significantly larger CPGE current compared to their thicker counterparts, and given the decrease in linear current with reduced thickness, we expect a *substantial enhancement in chiral sensitivity for ultrathin TI-based sensors*. Our calculations for the circular response, combined with available experimental data for the linear response, suggest that chiral sensitivity of below 1% is within reach.

**Performance-optimized incident light**

Up to this point, we have already established ultrathin TI films as advanced optical chirality detectors, demonstrating the enhancement of circular photogalvanic current by reducing film thickness to the nanometer scale. Here, we delve deeper into this phenomenon, investigating the influence of various parameters of incident light on the resulting helicity-dependent current within these ultrathin films. Our computational results, presented in Figure 5, reveal that by carefully tuning the frequency ($\omega$), angle of incidence ($\theta$), intensity ($A_0$), and delay time ($\tau$) of the incident light, we can significantly amplify the circular photogalvanic current ($C$). Our calculations indicate a direct relationship between decreasing the frequency of the incident light and an increase in the circular photogalvanic current. However, this enhancement is constrained by the hybridization gap, a characteristic energy gap that emerges in ultrathin TI films due to the overlap of surface states [37, 38]. For the circular photogalvanic effect to occur, the photon energy must exceed the hybridization gap [39], which is reported to be 252, 138, 70, and 41 meV for film thicknesses of 2-5QLs, respectively [37]. Moreover, excessively high frequencies can induce bulk-like excited states, limiting the frequency range to interband transitions within the Dirac cone [25].

Figure 5(b) shows the dependence of the circular photogalvanic current on the angle of incidence ($\theta$, cf. Figure 2). While increasing $\theta$ generally enhances the current, this trend is mitigated for angles exceeding about 60°, as predicted by Eq. 2 and seen in Figure 3(a). Additionally, the influence of light intensity on the circular photogalvanic current is explored in Figure 5(c). As logically expected, higher light intensities correlate with stronger currents. Lastly, Figure 5(d) presents the dependence of the current on delay time ($\tau$). Although increasing the delay time can amplify the HDPC, to avoid complications arising from photothermoelectric effects, the delay time should be confined below 100 picosecond range. In summary, our findings highlight the efficacy of tuning the incident light to enhance the resulting circular photogalvanic current in ultrathin TI films. By systematically reducing the frequency, increasing the angle of incidence, augmenting the light intensity, and optimizing the delay time, one can achieve substantial improvements in the circular response current. These results therefore offer valuable insights towards further development of existing and novel optoelectronic devices based on TI thin films.

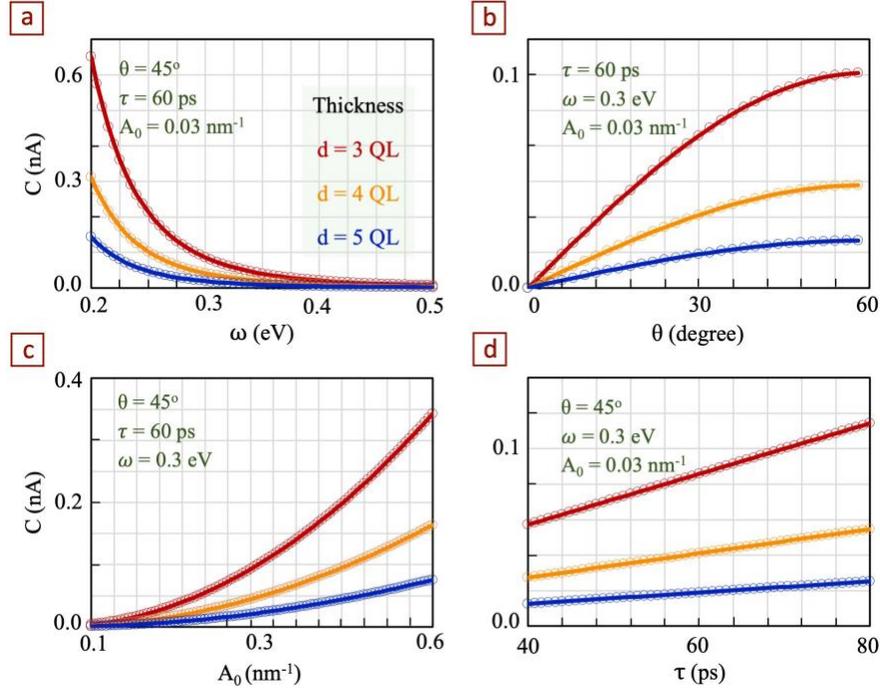

FIG. 5. Dependence of the circular photogalvanic current ($C$) in an ultrathin TI film (of irradiated area $S$ = 1.0 $\mu m^2$) on the parameters of incident light. (a) Frequency dependence, showing a direct correlation between lower frequencies and higher currents, constrained by the hybridization gap. (b) Angle of incidence dependence, revealing an increase in current with increasing $\theta$ up to 60°. (c) Intensity dependence, demonstrating a positive correlation between higher intensities and stronger currents. (d) Delay time dependence, highlighting the enhancement of the HDPC with increasing delay time, subject to limitations imposed by photothermoelectric effects.

**Straintronic enhancement of chiral response**

Strain engineering, the intentional application of mechanical stress to modify material properties, is a wellestablished routine for tuning the band structure and functional properties of ultrathin layered materials, including the TI films. In thin films, strain is known to significantly modulates the surface states in $Bi_2Se_3$ family [40], as experimentally observed near the grain boundaries [41], or in proximity to lattice sites where heavier elements like Te have been substituted, hence it presents an effective approach for engineering the band gap and manipulating Dirac states in these TI systems. Here, we demonstrate that by exerting appropriate lateral strain, one can efficiently tune the band structure and the topological state of ultrathin TI films towards significantly enhancing HDPC and their chiral sensitivity.

Strain is defined as the change in the displacement vector $U$ relative to the original, unstrained configuration. It can be mathematically represented by the following strain tensor [38]:

$$\varepsilon_{ij} = \frac{1}{2}\left(\frac{\partial U_i}{\partial x_j} + \frac{\partial U_j}{\partial x_i}\right) \tag{5}$$

The in-plane biaxial strain ($\epsilon_\parallel$) is quantitatively defined as: ($\epsilon_\parallel = \frac{a-a_0}{a_0}$), where $a$ represents the strained lattice parameter and $a_0$ denotes the unstrained lattice parameter in the $x$–$y$ plane. The application of biaxial strain induces changes in the lattice parameter along the $z$-direction as well, which consequently affects the band gap at the $\Gamma$ point. This relationship between in-plane and out-of-plane strain is characterized by the Poisson ratio.

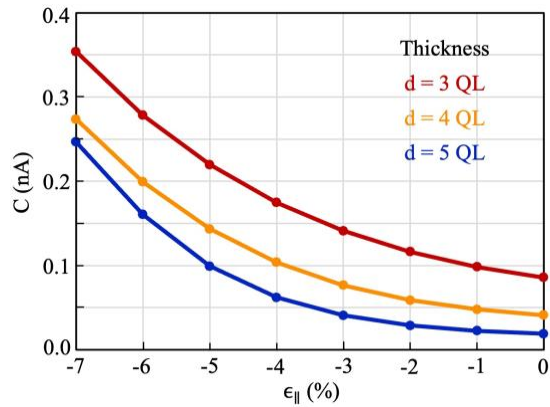

FIG. 6. Calculated HDPC in strained ultrathin TI films, of thickness 3-5QL. The results demonstrate a multifold enhancement of the circular photogalvanic current with compressive biaxial strain, highlighting the potential for straintronic control of chiral sensitivity.

In $Bi_2Se_3$, the Poisson ratio of 0.27 has been established, consistent with previous theoretical studies[42, 43]. The application of biaxial strain modifies the interatomic distances, consequently influencing the overlap of surface state wave functions. Compressive biaxial strain enhances the overlap between the top and bottom surface state wave functions in a thin-film TI, resulting in a larger hybridization gap [38]. Conversely, tensile strain diminishes the said overlap, reducing the hybridization gap, and a strain of 8% is sufficient to entirely close the gap for $N_{QL} > 3$. The enhanced hybridization gap due to compressive strain results in a reduced density of states near the Dirac point and modified optical selection rules. Consequently, we anticipate an enhancement in the circular photogalvanic effect due to the favorable tuning of the Berry curvature [25].

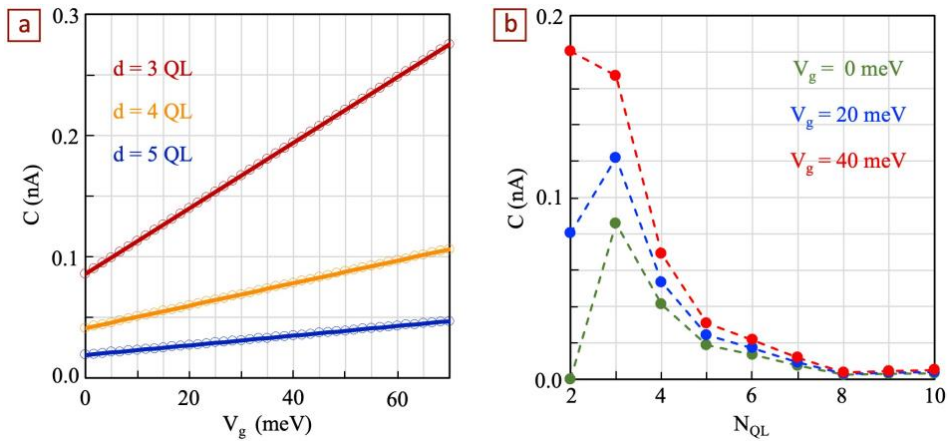

FIG. 7. Gate voltage dependence of the circular photogalvanic current in ultrathin TI films. (a) Variation of circular response current as a function of gate voltage for films of thickness 3, 4, and 5 QL. (b) Thickness dependence of the circular response current for gate voltages of 0, 20, and 40 meV. One should notice the enhancement of the circular photogalvanic current through gate voltage modulation, and the breaking of inversion symmetry in the gated 2QL sample. Here, the irradiated area of all samples was $S = 1.0$ $\mu m^2$, and the parameters of incident light were same as in Figure 4 ($\theta = \pi/4$, $\omega = 0.3$ eV, $A_0 = 0.03$ nm$^{-1}$, and $\tau = 60$ ps).

Our calculations for samples with thicknesses 3, 4 and 5QL, shown in Fig. 6, indeed demonstrate that compressive biaxial strain ($\epsilon_\parallel < 0$) significantly increases the HDPC in ultrathin TI film. We thereby reveal that controlled strain application can indeed serve as an efficient method for modulating and optimizing the optical response of ultrathin TI films, offering a viable approach for optimization of the chiral response in particular.

**Gate-tuned performance**

Last but not least, we investigate the impact of gate voltage on the circular photogalvanic current in ultrathin TI films. While the combined application of gate voltage and bias has been previously employed to distinguish linear responses from HDPC [13, 14], our focus here lies in exploring the potential of gate voltage to enhance the HDPC. As previously discussed, in thick TI films, the application of a gate voltage can induce a shift in the chemical potential towards the bulk bands, thereby increasing the scattering rate out of the topological surface states and consequently diminishing the circular photoresponse. Conversely, ultrathin TI films, owing to their reduced thickness and the corresponding decrease in the bulk-to-surface ratio, are less susceptible to this issue. In these ultrathin films, gating can be employed as a tool to controllably enhance the CPGE current and improve the performance of optical chirality detection.

Figure 7(a) presents our results, showing the calculated circular photogalvanic current as a function of out-of-plane applied gate voltage ($V_g$), for samples with thickness from 3 to 5 QLs. Moreover, in Figure 7(b), we compare the thickness dependence of the same current for specific gate voltage values of 0, 20, and 40 meV. Our findings demonstrate that the application of gate voltage can indeed effectively amplify the circular photogalvanic current in TIs. Furthermore, we find that imposing a gate voltage on a 2QL sample breaks inversion symmetry, giving rise to a non-zero circular photogalvanic current, that can exceed the previously found maximum for the 3QL thickness of the film for $V_g$ > 35 meV (for other parameters as given in Figure 7). This is in stark contrast to the zero current observed in the absence of a gate voltage due to the inherent inversion symmetry of the 2QL sample. This proves that gate voltage can also serve as a powerful tool for enhancing HDPC, thereby further improving the performance of optical chirality sensors based on ultrathin TIs.

## SUMMARY AND CONCLUSION

In summary, after having developed a method to isolate the helicity-dependent photocurrent response to incident light, a crucial quantity for optical chirality detection, we clearly demonstrated the unrivaled potential of ultrathin topological insulators (TIs) as a platform for miniaturized yet ultrasensitive optical chirality sensors. Next to the inherent chiral response of the TIs, such devices would directly benefit from suppression of the unwanted bulk effects, the structural inversion asymmetry, and the strongly favorable thickness dependence of the electronic properties and the chiral optical response in the thin limit of the material.

Moreover, we have revealed that the latter advanced response to chiral light can be further enhanced by optimally focused laser beam and the parameters of the incident light, and can also be broadly tuned by straining the sensor and/or back gating. Combining these optimization strategies, our calculations suggest such an enhancement in helicity-dependent photocurrent that is bound to translate into chiral sensitivity below 1%, significantly surpassing the incumbent technology.

It is worth noting that, in TIs that exhibit a relatively large hexagonal warping (such as $Bi_2Te_3$), an additional contribution to the helicity-dependent current arises when exposed to circularly polarized light [44]. This contribution is proportional to the out-of-plane spin texture and remains non-zero even for orthogonal incidence. Consequently, this effect can be exploited to further enhance the performance of optical chirality sensors and can be used as an additional booster to the chiral sensitivity of ultrathin TIs.

With the above, we earnestly advocate for the use of ultrathin TI films as an advanced platform for compact yet highly performant optical chirality detectors, reaching far beyond the present state-of-the-art. Our findings directly pave the way for the development of miniaturized sensors with a potential to revolutionize chiral analysis in various scientific domains, which will find broad applications in e.g. biomedical diagnostics, drug discovery, materials characterization, and emergent opto-electronic and other technologies.

## AUTHOR CONTRIBUTIONS

Mohammad Shafiei: Conceptualization, methodology, programming, editing. Sahar Safavi Moayeri: Conceptualization, programming, writing – original draft. Milorad V. Milošević: Conceptualization, supervision, writing – review and editing.

## ACKNOWLEDGMENTS

This research was supported by the Research Foundation-Flanders (FWO-Vlaanderen), the Special Research Funds (BOF) of the University of Antwerp, and the FWO-FNRS EoS-ShapeME project.

## DECLARATION OF COMPETING INTEREST

The authors declare that they have no known competing financial interests or personal relationships that could have appeared to influence the work reported in this paper.

* milorad.milosevic@uantwerpen.be